\documentstyle[aps,prl,psfig,multicol]{revtex}
\begin{document}
\preprint{\today}
\title{Nonergodicity parameters for a molecular liquid: \\
A comparison between mode coupling theory and simulation}
\author{Christoph Theis and Rolf Schilling}
\address{Institut f\"ur Physik, Johannes Gutenberg--Universit\"at,
Staudinger Weg 7, D--55099 Mainz, Germany}
\maketitle

\begin{abstract}
We apply the mode coupling theory (MCT) which was recently worked out
for molecular liquids to a liquid of diatomic, rigid molecules. Making
use of the static correlators $S_{l l^\prime}^m(q)$ from a molecular 
dynamical simulation, we have solved the MCT--equations for the 
nonergodicity parameters $f_l^m(q) = \lim_{t \to\infty} S_{l l}^m(q,t)/
S_{l l}^m(q)$, assuming that all correlators are nonzero for 
$l = l^\prime = 0,1,...,l_{co}$, only. Depending on $l_{co} = 0,1,2$
we obtain different types of ideal glass transitions with transition
temperatures $T^{(l)}_c$ which are below $T^{MD}_c$ as deduced from 
the MD--simulation. The q--dependence of the critical nonergodicity 
parameter $(f_l^m)_c^{MD}$ from the simulation is reasonably well 
reproduced by the corresponding MCT--result. The influence of  
a strong and weak diagonalization approximation  on the results is
discussed.
\end{abstract}


\begin{multicols}{2}

\section{Introduction}
\label{sec1}

The use of mode coupling theory (MCT) is one of the most important 
attempts to describe the {\em structural} glass transition. Restricted
to {\em simple} liquids, it has provided an equation of motion for the 
normalized density correlator
\begin{equation}
\label{eq1}
\Phi(q,t) = \frac{S(q,t)}{S(q)}
\end{equation}
where $S(q)$ is the corresponding static correlation function. MCT, in
its {\em idealized} version, interprets the glass transition as a
dynamical transition at a critical temperature $T_c$ (or critical
density $n_c$) from a ergodic (supercooled liquid) to a nonergodic
phase (glass). In the latter phase the density $\rho(q,t)$ does not
approach its equilibrium value $\rho_0 = \langle \rho(q,t) \rangle$ in
the long time limit, i.e. the correlation function $\Phi(q,t)$ of the
density fluctuations $\delta\rho(q,t) = \rho(q,t) - \rho_0$ does not
decay to zero for $t \to\infty$. Therefor one can use the
nonergodicity parameter (NEP)
\begin{equation}
\label{eq2}
f(q) = \lim_{t \to\infty} \Phi(q,t)
\end{equation}
as an order parameter for the glass transition. $f(q)$ vanishes above
and becomes nonzero at $T_c$. In the long time limit $t \gg t_0$ the
MCT--equations involve only $S(q)$ as an input quantity. Here, $t_0$
is a microscopic time of order $10^{-13}$ sec. This fact demonstrates
the {\em microscopic} nature of MCT, because $S(q)$ can be calculated
from the potential energy $V({\it\bf x}_1,...,{\it\bf x}_N)$ which
only depends on the microscopic configuration \{ ${\it\bf x}_n$ \} of
the particles. The discussion of the MCT--equations for T in the
vicinity of $T_c$ leads to several interesting predictions for the
time and temperature dependence of $\Phi(q,t)$. However, due to the
existence of ergodicity restoring processes in real glass formers
below $T_c$, which are taken into account by the {\em extended} MCT,
the transition is smeared out. Nevertheless, a couple of experimental
and numerical investigations have clearly shown the validity of
several predictions of idealized MCT. For more details the reader may
consult the reviews~\cite{gotze91,gotze92,kim90,schilling94,kob97,yip95}.

These tests of idealized MCT (simple liquids) were mainly {\em
qualitative}, i.e. the validity of the asymptotic power laws for the
$\alpha$-- and $\beta$--relaxation regimes was studied. Only very few
{\em quantitative} test exist, which is, besides the calculation of
the critical temperature (or critical density), the q--dependence of
the critical NEP $f^c(q)$ at $T_c$ (or $n_c$) and the critical
amplitude $h(q)$. This was done for a monoatomic~\cite{bengtzelius86}
(only $T_c$ was determined) and binary Lennard--Jones liquid
~\cite{nauroth97}, for a hard sphere~\cite{pusey90} and a soft sphere
system~\cite{gotze92,latz90} and for the molecular liquids
orthoterphenyl~\cite{petry91} and water~\cite{sciortino97}. Although
the two latter liquids contain orientational degrees of freedom
(ODOF), the analysis of the experimental~\cite{petry91} and numerical
results~\cite{sciortino97} was done in the framework of MCT for {\em
simple} liquids, which does not account for the ODOF. A first attempt
to include ODOF and their coupling to the translational degrees of
freedom (TDOF) has recently been made for water~\cite{fabbian97}.
These authors suggest a semi--schematic model where the full
q--dependence is considered, but the influence of the ODOF is
condensed in a single parameter $\chi_R$. The phenomenological
introduction of $\chi_R$ can be viewed as a ``renormalisation'' of the
mode coupling strength for the TDOF (center of mass). 

Because most of the glass formers are of molecular origin, it is
important to extend MCT to {\em molecular} liquids. This has recently
been done for a single dumb-bell molecule in an isotropic
liquid~\cite{franosch97}, for linear molecules~\cite{schilling97} and
arbitrary molecules~\cite{schmitz97,kawasaki97,theis97}. 
In the present paper we will
perform a quantitative test of the molecular MCT for a liquid of rigid
diatomic molecules by comparing the corresponding nonergodicity
parameters as obtained from a recent MD--simulation with those from
molecular MCT.

The outline of our paper is as follows. The next section contains a
descripition of the molecular MCT--equations for linear molecules. The
diatomic liquid and the computational details are described in section
3. Section 4 contains the comparison between MCT and numerical results
and the final section presents a summary and some conclusions.

\section{Molecular MCT}
\label{sec2}

In this section we give a short summary of the application of MCT to
molecular liquids. We will restrict ourselves to linear molecules.
For details the reader is referred to Ref.~\cite{schilling97}. The
starting point is the tensorial one--particle density
\begin{eqnarray}
\label{eq3}
& &
\rho_{l m}(\mbox{\boldmath $q$},t) = \sqrt{4\pi} \; i^l \sum_{n=1}^{N}
\exp(i \mbox{\boldmath $q \; x$}_n(t)) \; 
Y_{l m}(\Omega_n(t)) \, , \nonumber \\
& & l = 0,1,2,... \quad, -l \le m \le l
\end{eqnarray}
where $\mbox{\boldmath $x$}_n(t)$ and $\Omega_n(t) = ( \theta_n(t), \phi_n(t) )$
are, respectively, the center of mass position and the Euler angles of
the n--th molecule at time t. Due to the axial symmetry of the
molecule the third Euler angle $\chi_n(t)$ does not play a role. N is
the number of molecules and $Y_{l m}$ are the spherical harmonics. The
factor in front of the sum is for technical convenience. The
generalization of the density correlator $S(q,t)$ (cf. eq.(\ref{eq1}))
to molecular liquids of linear molecules is straightforward:
\begin{equation}
\label{eq4}
S_{l m, l^\prime m^\prime}(\mbox{\boldmath $q$},t) = \langle \rho_{l
m}^\ast(\mbox{\boldmath $q$},t) \rho_{l^\prime m^\prime}
(\mbox{\boldmath $q$},0) \rangle
\end{equation}
where $\ast$ denotes complex conjugation and the angular brackets
refer to the canonical average over the initial conditions in phase
space. Generally, the correlation matrix $\mbox{\boldmath $\sf S$}
(\mbox{\boldmath $q$},t)
\equiv ( S_{l m, l^\prime m^\prime}(\mbox{\boldmath $q$},t) )$ 
is nondiagonal in
both $l$ and $m$. In the q--frame, i.e. for $\mbox{\boldmath $q$} = 
\mbox{\boldmath $q$}_0
\equiv (0,0,q)$, $q = |\mbox{\boldmath $q$}|$, \boldmath ${\sf S}
(q,\mbox{\unboldmath $t$})$ \unboldmath becomes
diagonal in $m$:
\begin{equation}
\label{eq5}
S_{l m, l^\prime m^\prime}(\mbox{\boldmath $q$}_0,t) \equiv S_{l
l^\prime}^m(q,t) \delta_{m m^\prime}
\end{equation}
The matrix elements $S_{l l^\prime}^m(q,t)$ are real, and nonzero for
$0 \le |m| \le \min({l,l^\prime})$, only. In addition they
fulfill~\cite{schilling97}
\begin{equation}
\label{eq6}
S_{l l^\prime}^{-m}(q,t) = S_{l l^\prime}^m(q,t) .
\end{equation}
In the third and fourth section we will restrict ourselves to those
correlators with $l=l^\prime=0,1$ and 2. In that case, only six
independent correlators $S_l^m(q,t) \equiv S_{l l}^m(q,t)$, $0 \le m
\le l$ exist.

A closed set of equations can be derived for the Laplace transform
$\hat{\sf S}(\mbox{\boldmath $q$},z)$ of \boldmath${\sf S}(q,
\mbox{\unboldmath $t$})$\unboldmath by use of the Zwanzig--Mori
projection formalism in combination with the mode coupling
approximation which replaces the slow part of the time dependent
memory kernel by a linear combination of bilinear terms $S_{l_1 m_1,
l_1^\prime m_1^\prime}(\mbox{\boldmath $q$}_1,t) S_{l_2 m_2,l_2^\prime
m_2^\prime}(\mbox{\boldmath $q$}_2,t)$. The final result for 
the MCT--equations is as
follows~\cite{schilling97}:
\begin{equation}
\label{eq8}
\hat{\sf S}(\mbox{\boldmath $q$},z) = - \left[ z {\sf 1} + \hat{\sf
K}(\mbox{\boldmath $q$},z) {\sf S}^{-1}(\mbox{\boldmath $q$}) 
\right]^{-1} {\sf
S}(\mbox{\boldmath $q$})
\end{equation}
with
\begin{equation}
\label{eq9}
\hat{K}_{l m,l^\prime m^\prime}(\mbox{\boldmath $q$},z) = \sum_{\alpha
\alpha^\prime} q_l^\alpha(\mbox{\boldmath $q$}) 
q_{l^\prime}^{\alpha^\prime}
(\mbox{\boldmath $q$}) \hat{k}_{l m,l^\prime m^\prime}^{\alpha \alpha^\prime}
(\mbox{\boldmath $q$},z)
\end{equation}
\begin{equation}
\label{eq10}
q_l^\alpha(\mbox{\boldmath $q$}) = \left\{
\begin{array}{c@{\quad,\quad \alpha=}c}
q & T \\ \sqrt{l(l+1)} & R
\end{array} \right.
\end{equation}
and
\begin{equation}
\hat{\sf k}(\mbox{\boldmath $q$},z) = - \left[ z {\sf J}^{-1} + {\sf
J}^{-1} \hat{\sf M}(\mbox{\boldmath $q$},z) {\sf J}^{-1} \right]^{-1} .
\end{equation}
${\sf J} \equiv \left( J_{l m,l^\prime m^\prime}^{\alpha
\alpha^\prime} \right)$ is the correlation matrix of the translational
($\alpha = T$) and rotational ($\alpha = R$) current density $j_{l
m}^\alpha(\mbox{\boldmath $q$})$~\cite{schilling97}. 
The kernel ${\sf J}^{-1}{\sf M}(\mbox{\boldmath
$q$},t){\sf J}^{-1}$ cannot be calculated exactly. 
The mode coupling approximation
yields for its slow part (which determines the long time relaxation):
\begin{eqnarray}
\label{eq12}
& & \left[ \left( {\sf J}^{-1} {\sf M}(\mbox{\boldmath $q$},t) 
{\sf J}^{-1}
\right)^{\alpha \alpha^\prime}_{\lambda \lambda^\prime}
\right]_{\mbox{slow}} \approx m^{\alpha \alpha^\prime}_{\lambda
\lambda^\prime}(\mbox{\boldmath $q$},t) = \nonumber 
\\ & & = \frac{1}{2N} \sum_{\mbox{\boldmath $q$}_1 
\mbox{\boldmath $q$}_2} \sum_{\lambda_1
\lambda_2 \lambda_1^\prime \lambda_2^\prime} V^{\alpha \alpha^\prime}
(\mbox{\boldmath $q$} \lambda \lambda^\prime | \mbox{\boldmath $q$}_1 
\lambda_1 
\lambda_1^\prime ; \mbox{\boldmath $q$}_2 \lambda_2 \lambda_2^\prime)  
\times \nonumber \\
& & \: \times S_{\lambda_1 \lambda_1^\prime}(\mbox{\boldmath $q$}_1,t) 
S_{\lambda_2
\lambda_2^\prime}(\mbox{\boldmath $q$}_2,t) .
\end{eqnarray}
Here we have used the short hand notation $\lambda = (l, m)$. The
vertices $V^{\alpha \alpha^\prime}$ only depend on the static
correlators $S_{\lambda \lambda^\prime}(\mbox{\boldmath $q$})$ 
or equivalently
on the direct correlations functions $c_{\lambda
\lambda^\prime}(\mbox{\boldmath $q$})$. The explicit expression for $V^{\alpha
\alpha^\prime}$ can be found in Ref.~\cite{schilling97}.
Eqs.(\ref{eq8})--(\ref{eq12}) are the closed set of MCT--equations
for linear molecules, and their structure is quite similar to that of
the MCT--equations for simple liquids. But due to the presence of ODOF
and their coupling to TDOF, which is in the vertex functions
$V^{\alpha \alpha^\prime}$, the mode coupling polynomial (\ref{eq12}) is
much more involved. Therefore, in a first step we will simplify these
equations by assuming diagonality of ${\sf S}(\mbox{\boldmath $q$},t)$ 
and
${\sf m}(\mbox{\boldmath $q$},t)$ with respect to $l$. 
We stress that this diagonalization approximation is also assumed for 
the static correlators, in order to keep the coupled set of equations
as simple as possible.
Since ${\sf J}$ is
already diagonal, eqs.(\ref{eq8})--(\ref{eq12}) allow for a self
consistent solution for the normalized correlators
\begin{equation}
\label{eq13}
\Phi_l^m(q,t) = \frac{S_{l l}^m(q,t)}{S_{l l}^m(q)}
\end{equation}
in the q--frame. Then, the corresponding nonergodicity parameters
\begin{equation}
\label{eq14}
f_l^m(q) = \lim_{t \to\infty} \Phi_l^m(q,t) = - \lim_{z \to 0}
z \hat{\Phi}_l^m(q,z)
\end{equation}
are solutions of the following set of coupled equations:
\begin{equation}
\label{eq15}
f_l^m(q) = \left[ 1 + \sum_{\alpha \alpha^\prime} q_l^\alpha(q)
q_{l^\prime}^{\alpha^\prime}(q) \left( \left[ {\sf
F}_l^m(q;\{f_l^m(q)\}) \right]^{-1} \right)^{\alpha \alpha^\prime}
\right]^{-1}
\end{equation}
with the 2x2 matrix
\begin{equation}
\label{eq16}
{\sf F}_l^m = \left( 
\begin{array}{cc}
(F_l^m)^{TT} & (F_l^m)^{TR} \\
(F_l^m)^{RT} & (F_l^m)^{RR}
\end{array} \right) .
\end{equation}
In the thermodynamic limit it is:
\begin{eqnarray}
\label{eq17}
& & \left( F_l^m(q;\{f_l^m(q)\}) \right)^{\alpha \alpha^\prime} =
\nonumber \\ & & = 
\int_0^\infty dq_1 \int_{|q-q_1|}^{q+q_1} dq_2 \left(
{\sf H}_l^m(q, q_1, q_2;\{f_l^m(q)\} ) \right)^{\alpha \alpha^\prime}
\end{eqnarray}
where
\begin{eqnarray}
\label{eq18}
& & \left( {\sf H}_l^m(q, q_1, q_2;\{f_l^m(q)\}) 
\right)^{\alpha \alpha^\prime}
= \nonumber \\ & & = 
\sum_{l_1 l_2}{}^\prime \sum_{m_1 m_2} v^{\alpha \alpha^\prime}(q l m|
q_1 l_1 m_1; q_2 l_2 m_2) f_{l_1}^{m_1}(q_1) f_{l_2}^{m_2}(q_2) .
\end{eqnarray}
$\sum_{l_1 l_2}^\prime$ denotes the summation over $l_1$ and $l_2$
such that
\begin{equation}
\label{eq19}
\begin{array}{cc}
l_1+l_2+l \quad \mbox{is even} \\
|l_1 - l_2| \le l \le l_1 + l_2.
\end{array}
\end{equation}
This restriction results from the Clebsch--Gordon coefficients $C(l_1
l_2 l; 0 0 0)$ which enter into $v^{\alpha \alpha^\prime}$. 
The expressions for the $v^{\alpha \alpha^\prime}$ are rather
involved. Therefore we do not give them explicitely, here. They can be
deduced from $V^{\alpha \alpha^\prime}$ (see second paper
of ref.~\cite{schilling97}) by use of: 
\newline\noindent i) transformation to the
q-frame where one has to use the transformation properties of the
correlators under rotations (see again second paper of
ref.~\cite{schilling97}) 
\newline\noindent ii) the diagonalization approximation with respect
to l and
\newline\noindent iii) the thermodynamic limit $N \to\infty, V
\to\infty$ such that $N/V = const$, which replaces the sums over 
\boldmath $q_1, q_2$\unboldmath in eq.(\ref{eq12}) by integrals; then
taking into account the conservation of momentum \boldmath $q = q_1 +
q_2$ \unboldmath and performing the angular integration one obtains
the integral representation of eq.(\ref{eq17})

$V^{\alpha \alpha^\prime}$. This type of approximation will be called
{\em weak} diagonalization approximation (WDA). Assuming in addition
the diagonality of $v^{\alpha \alpha^\prime}$ with respect to
$\alpha$, and accordingly the diagonality of ${\bf\sf F}_l^m$ (which
we call the {\em strong} diagonalization approximation (SDA)), the
eqs.(\ref{eq15})--(\ref{eq18}) simplify even more. It is the SDA which
has been used for the calculation of $f_l^m(q)$ for a dumb-bell
molecule in a liquid of hard spheres~\cite{franosch97} and for a
system of dipolar hard spheres~\cite{schilling97}. 
The diagonalization
approximation for the time--dependent {\em and} the static correlators
with respect to the $l$'s implies (\ref{eq19}). This
restriction has two consequences. First, the eq.(\ref{eq15}) for the
$f_l^m$ with $l$ even may contain on its r.h.s. bilinear terms $f_l^m
f_l^m$, but for $l$ odd this cannot happen. Therefore $f_l^m$ with $l$ even
will exhibit a type--B transition and those with odd $l$ may show a
type--A or type--B transition. In case of a type--A transition the
corresponding transition temperature is below that of the type--B
transition for $f_l^m$ with $l$ even. Second, whereas eq.(\ref{eq15})
for $f_0^0$ contains on its r.h.s. $f_0^0 f_0^0$, $f_1^{m_1} f_1^{m_2}$, 
$f_2^{m_1} f_2^{m_2}$ etc., the equation for, e.g. $f_2^m$ involves
$f_1^{m_1} f_1^{m_2}$, $f_2^{m_1} f_2^{m_2}$, $f_0^0 f_2^{m_2}$ etc.,
but {\em not} $f_0^0 f_0^0$, due to the diagonalization approximation
of the {\em static} correlators. Therefore the freezing of the center of
mass order parameter $f_0^0$ does not necessarily imply the freezing
of $f_2^m$.  $f_2^m$ may freeze at a lower temperature via a type--B
transition. Without the diagonalization approximation with respect to
the $l$'s, we expect {\em all} $f_{l l^\prime}^m(q)$ to freeze 
via a type-B
transition at a {\em single} critical temperature
$T_c$. For more details on type--A and type--B
transitions the reader is referred to the schematic models which are
discussed in ref.~\cite{gotze91} and in ref.~\cite{franosch94}.

\section{Model}
\label{sec3}

In order to test the MCT--predictions, a 
MD--simulation for a system of diatomic, rigid 
molecules~\cite{kammererI,kammererII,kammererIII} was performed.
Each molecule is
composed of two different Lennard--Jones particles A and B, which are
separated by a distance d. Both particles have the same mass m. The
interaction between two molecules is given by the sum of the
interactions between the four particles which are given by the
Lennard--Jones potentials $V_{\alpha \beta}(r) = 4 \epsilon_{\alpha
\beta} \left[ (\sigma_{\alpha \beta} / r)^{12} - (\sigma_{\alpha
\beta} / r)^6 \right]$. The Lennard--Jones parameters are given by 
$\sigma_{AA} = \sigma_{AB} = 1$, $\sigma_{BB} = 0.95$, $\epsilon_{AA} =
\epsilon_{AB} = 1$ and $\epsilon_{BB} = 0.8$. We use lengths in units
of $\sigma_{AA}$ and $\epsilon_{AA}$ is taken as unit for the
temperature ($k_B = 1$). In these units we have found that $d = 0.5$
is an appropriate value in order to avoid formation of a liquid
crystalline phase. The MD--simulation was done for $N = 500$ molecules
in an (N,p,T)--ensemble with $p = 1.0$ and T between 0.477 and 5.0 by
use of the rattle algorithm. The length of the runs was long enough in
order to prepare a well equilibrated system. In order to increase the
statistics, average over at least eight independent runs was
performed. One of the main results we have found is the existence of a
type--B transition with a {\em single} transition temperature
$T_c^{MD} \cong 0.475$, at which the dynamics crosses over from
ergodic to quasi--nonergodic behavior. This temperature has been
deduced from the $\alpha$--relaxation time of the purely orientational
(i.e. ${\bf\it q} = 0$ in eq.~\ref{eq4})~\cite{kammererI}, purely
translational (i.e. $l = l^\prime = 0$ in
eq.(\ref{eq4}))~\cite{kammererII} and of the general correlators where
$q \not= 0$ and $l = l^\prime \not= 0$~\cite{kammererIII}. The
critical temperature $T_c$ can be located within about 2\% from these
data. For more details the reader is referred to
refs.~\cite{kammererI,kammererII,kammererIII}.

\section{Results}
\label{sec4}

As already mentioned in section \ref{sec2} we need for the solution of 
eqs.(\ref{eq15})--(\ref{eq18}) the static correlators $S_{l
l^\prime}^m(q)$ as a function of $(q,l,l^\prime,m)$ and temperature T.
We remind the reader that we use the q--frame representation, without
restricting generality. The MD--result for $S_{l l^\prime}^m(q)$ is
shown in Figure 1 for $T = 0.49$ and for $l = l^\prime = 0,1,2$. The
variation of $S_l^m(q) \equiv S_{l l}^m(q)$ with temperature is
smooth. From this Figure we observe that the q--dependence of
$S_l^m(q)$ strongly depends on $l$ and $m$. Whereas $S_0^0(q)$ and 
$S_1^0(q)$
resembles the behavior of $S(q)$ for a simple liquid, with e.g. a
well pronounced main maximum at $q_{max} \approx 2\pi$, the
q--dependence of the other correlators is qualitatively different. For
instance $S_2^1$ has its main peak at $q = 0$. 

\end{multicols}
\begin{figure}[hb]
\centerline{\psfig{figure=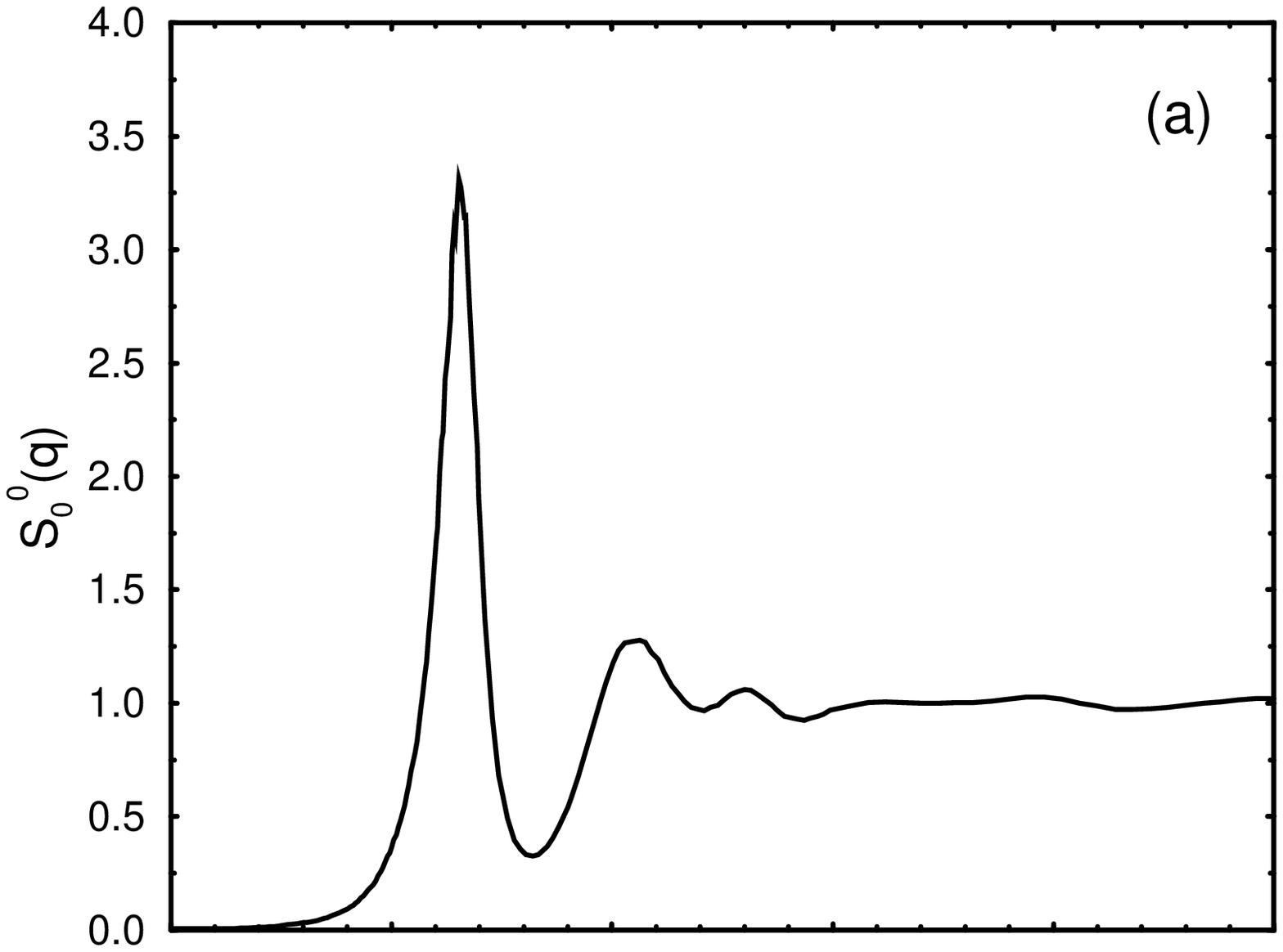,width=80mm} 
\psfig{figure=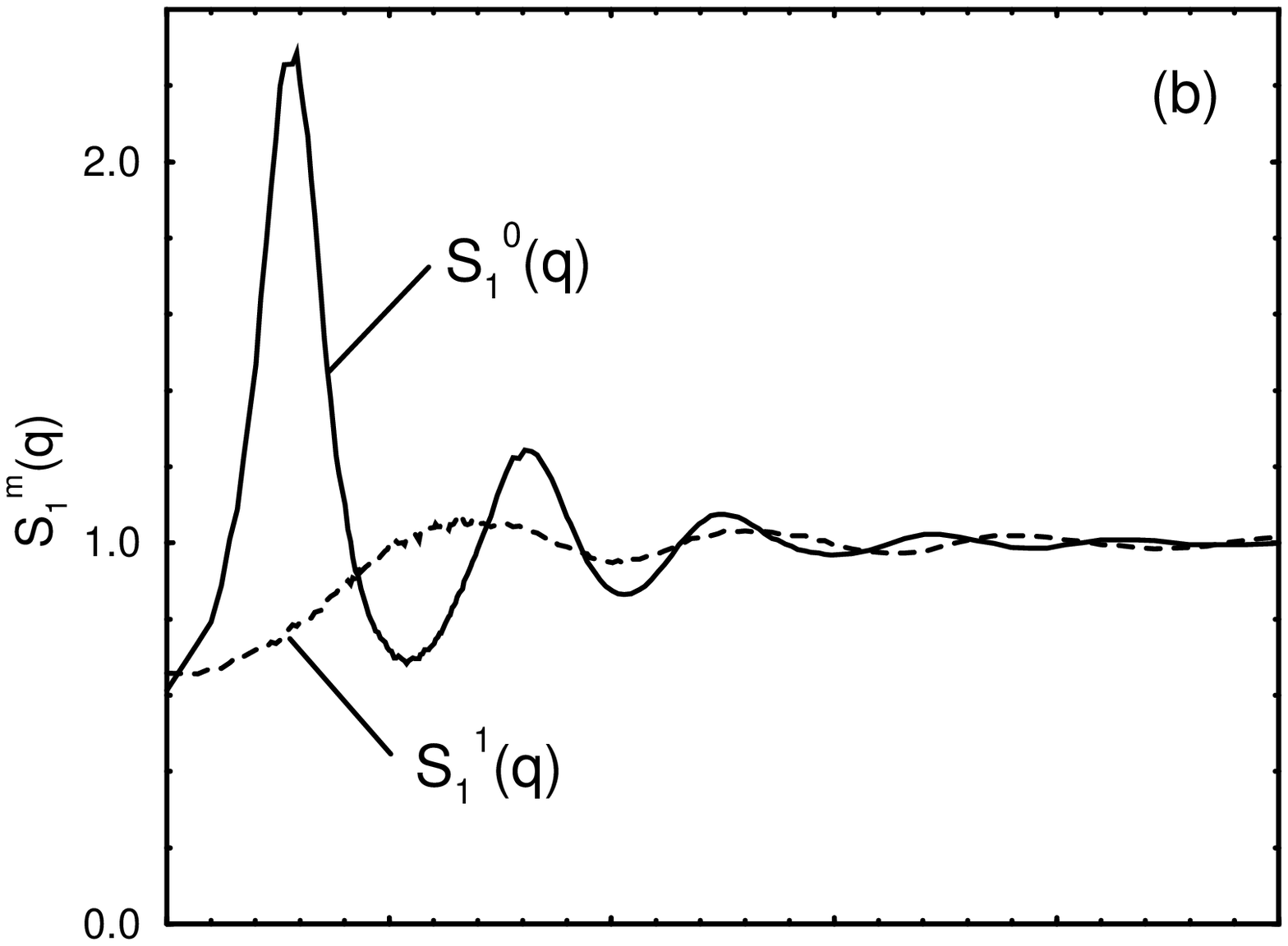,width=80mm} }
\centerline{\psfig{figure=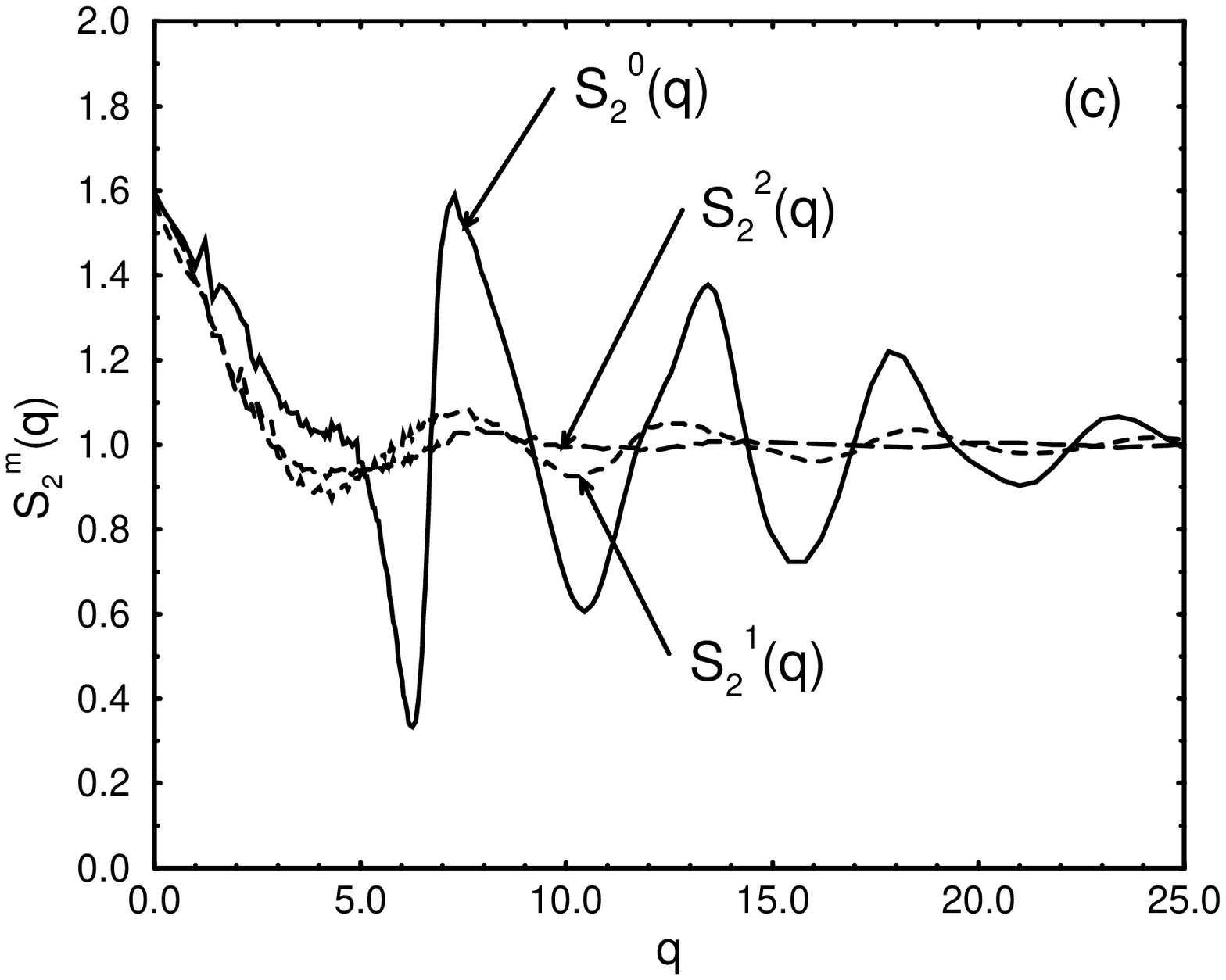,width=80mm}}
\caption[]{
$S_l^m(q)$ versus q for $T = 0.49$: (a) $l = m = 0$, (b) $l =
1, m = 0,1$ and (c) $l = 2,m = 0,1,2$. }
\end{figure}

\begin{multicols}{2}
As will be seen below,
we need $S_l^m(q)$ for temperatures below $T_{min} = 0.477$, the
lowest value of our MD--simulation. This has been achieved by a linear
extrapolation from the higher temperature regime ($T \ge 0.477$).
After smoothing the q--dependence of $S_l^m(q)$ by a spline under
tension, equations (\ref{eq15})--(\ref{eq18}) were solved by iteration.
This procedure is standard and had been used for simple liquids (see
e.g.~\cite{gotze91} and~\cite{nauroth97}). The iteration was stopped
after the maximum value of $\left| ( f_l^m(q) )^{(\nu+1)} - ( f_l^m(q)
)^{(\nu)} \right|$ was less than $10^{-6}$. We have used SDA and WDA (see
section \ref{sec2}). The $l$--dependence has been truncated at $l_{co}=0,1$ and
2. The integrations in (\ref{eq17}) were performed numerically. For
this the interval $[0,q_{co}]$ was divided into 300 {\em equidistant}
grid points. Since the MD--data were limited to $q \le q_{co} = 25$,
we were not able to check the sensitivity of the results for $f_l^m(q)$
on a variation with $q_{co}$. But there should not be a considerable
change by extending $q_{co}$ to higher values than 25, because
$S_l^m(q)$ already approaches one for $q = 25$ (for $S_{2 2}^0$ this is
not quite true) (cf. Fig.1).

Special attention must be paid to $q = 0$. Since $\left(
F_l^m(q,\{f_l^m(q)\}) \right)^{\alpha \alpha^\prime}$ for $l \not= 0$ is
proportional to $q^{-2}$ and $q^{-1}$ for $\alpha = \alpha^\prime = T$
and $\alpha \not= \alpha^\prime$, respectively, we have treated
eq. (\ref{eq15}) separately for $q = 0$ and $q \ge q_{min} \cong 0.08$.
The results for $f_l^m(q)$ in between were obtained by a polynomial
interpolation of degree four.

The first observation we make is that $f_l^m(q) \equiv 0$ for $T \ge
0.477$. This result, which holds for SDA, WDA and $l_{co} = 0,1,2$, is
unexpected, because for all known cases MCT overestimates the 
transition temperature
$T_c$. 
For instance, it is $T_c^{MCT}
\approx 2 T_c^{MD}$ for the binary Lennard--Jones
liquid~\cite{nauroth97}. For $T < T_c^{MD} \cong 0.475$ ideal glass
transitions are found for our molecular liquid. The transition
temperatures are given in Table 1. 
Here a comment is in order. The transition temperatures are determined 
whitin an accuracy of $10^{-3}$. Since the critical NEP vary as 
$(T-T_c)^\frac{1}{2}$ they are correct only within 3--5 \%. For an
accuracy of 1 \% and less one must determine $T_c$ more precisely.
This can be achieved by use of the T--dependence of the largest
eigenvalue $E_0(T)$ of the stability matrix~\cite{gotze91}, since
$(1-E_0(T))^2 \sim T_c-T$ for $T \le T_c$.

From this table we recognize that
the center of mass modes ($l = m = 0$) freeze at $T_c^{l=0} \cong
0.383$ independent on $l_{co}$ and SDA or WDA. This transition is of
type B. Ideal glass transitions for $l=1$ and 2 occur at lower
temperatures. Their transition scenario depends on whether $l_{co} =
1$ or $l_{co} = 2$. In case that $l_{co} = 1$, i.e. all $l \ge 2$ are
neglected, a type--A transition of $f_1^m(q)$ exists at $0.192$ for
WDA and at a slightly lower value of 0.187 for SDA. But for $l_{co} =
2$ both modes with $l = 1$ and $l = 2$ {\em simultaneously} freeze at
$T_c^{l=1} = T_c^{l=2} \cong 0.310 (0.256)$ for WDA (SDA). This
transition is of type B for $l = 2$ {\em and} $l = 1$. Whereas $l = 2$
{\em must} freeze via a discontinuous (type--B) transition, $l = 1$
could undergo a continuous transition (type--A). The coupling between
the modes with $l = 1$ and $l = 2$ in eq.(\ref{eq18}) is obviously
large enough to lead to a type--B transition for $l = 1$, too. These
possible scenarios are in accordance with the discussion at the end of
section \ref{sec2}. Since for WDA the r.h.s. of eq.(\ref{eq15})contains more
contributions from the coupling between ODOF and TDOF than for SDA, it
is reasonable that the critical temperatures for WDA are larger (or
equal) than those for SDA.

The NEP $\left(f_l^m\right)_c^{MD}$ obtained from the
MD--simulation~\cite{kammererIII} by a fit with the von Schweidler law
and $\left(f_l^m\right)^{MCT}$
from molecular MCT for $l_{co} =
2$, WDA and $T = T_c^{l=1} = T_c^{l=2} \cong 0.310$ are shown in
Figure 2. Whereas $\left(f_l^m\right)_c^{MD}$ for $l = 0,1,2$ and
$\left(f_l^m\right)_c^{MCT}$ for $l = 1,2$ are the {\em critical} NEP,
this is not true for $\left(f_0^0\right)^{MCT}$, since
$T_c^{l=0}>0.310$. Despite the approximations we made, the
q--dependence of $\left(f_l^m\right)_c^{MD}$ is reasonably well
reproduced by the MCT--result. Even the shoulder in
$\left(f_0^0\right)_c^{MD}$ at $q \approx 3$ is reproduced as a small
hump at that q--value. Apart from this prepeak, the q--variation of
$f_l^m$ is in phase with that of the corresponding static correlators
(cf. Fig.1). Whereas $\left(f_0^0\right)^{MCT} >
\left(f_0^0\right)_c^{MD}$ (as expected again, due to the mean field
character of MCT) the opposite is true for $f_l^m$ with $l=1$ and 2.
For SDA the result for $\left(f_l^m\right)^{MCT}$ at $T = T_c^{l=1} =
T_c^{l=2} \cong 0.256$ (not shown in Figure 2) is about 5 - 10 \%
below the corresponding result for WDA. 

\end{multicols}

\begin{table}
\begin{tabular}{|c||c|c|c|} \hline
& $T_c^{l=0}$ & $T_c^{l=1}$ & $T_c^{l=2}$ \\ \hline\hline
& 0.383 & & \\
\raisebox{1.5ex}[-1.5ex]{$l_{co} = 0$} & (0.383) & 
\raisebox{1.5ex}[-1.5ex]{---} & \raisebox{1.5ex}[-1.5ex]{---} \\ \hline
& 0.383 & 0.192 & \\
\raisebox{1.5ex}[-1.5ex]{$l_{co} = 1$} & (0.383) & (0.187) &
\raisebox{1.5ex}[-1.5ex]{---} \\ \hline
& 0.383 & 0.310 & 0.310 \\
\raisebox{1.5ex}[-1.5ex]{$l_{co} = 2$} & (0.383) & (0.256) & (0.256) \\
\hline
\end{tabular}
\noindent
\caption[]{Critical temperatures $T_c^l$ for $l_{co}=0,1,2$ and WDA. The
corresponding values for SDA are given in parantheses.
\protect\label{tab1}}
\end{table}

\begin{figure}[htb]
\centerline{\psfig{figure=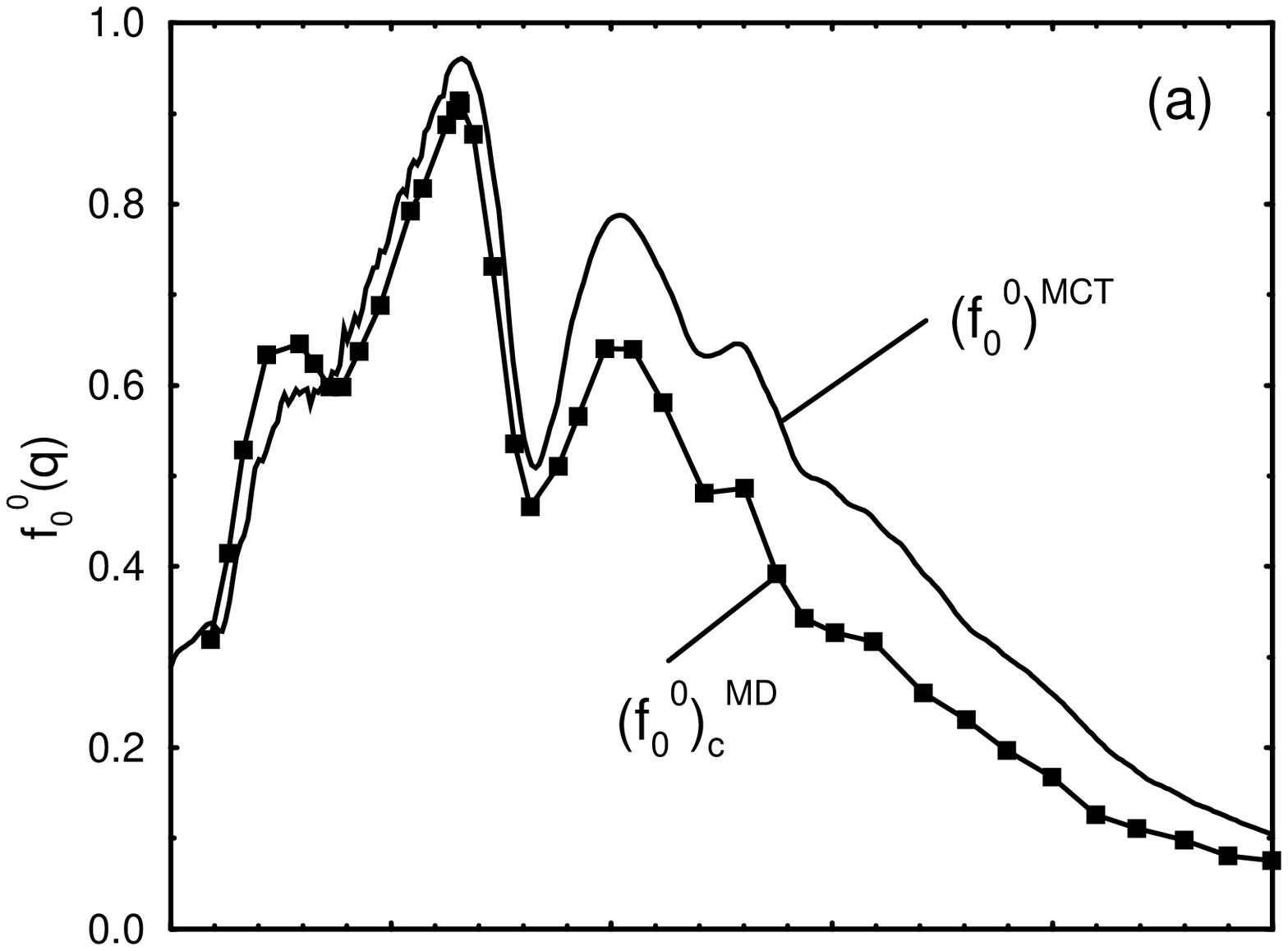,width=80mm}
\psfig{figure=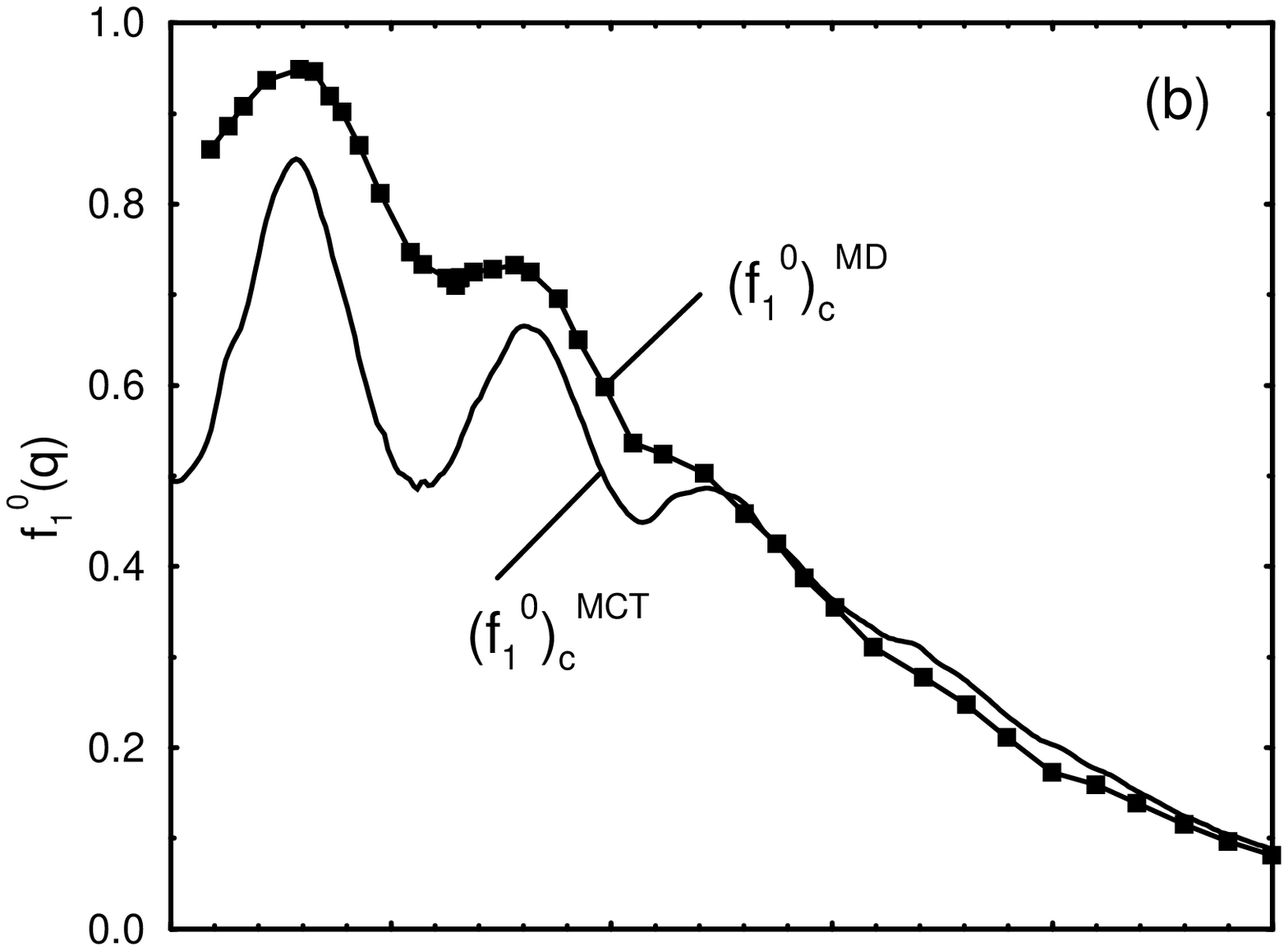,width=80mm} }
\centerline{\psfig{figure=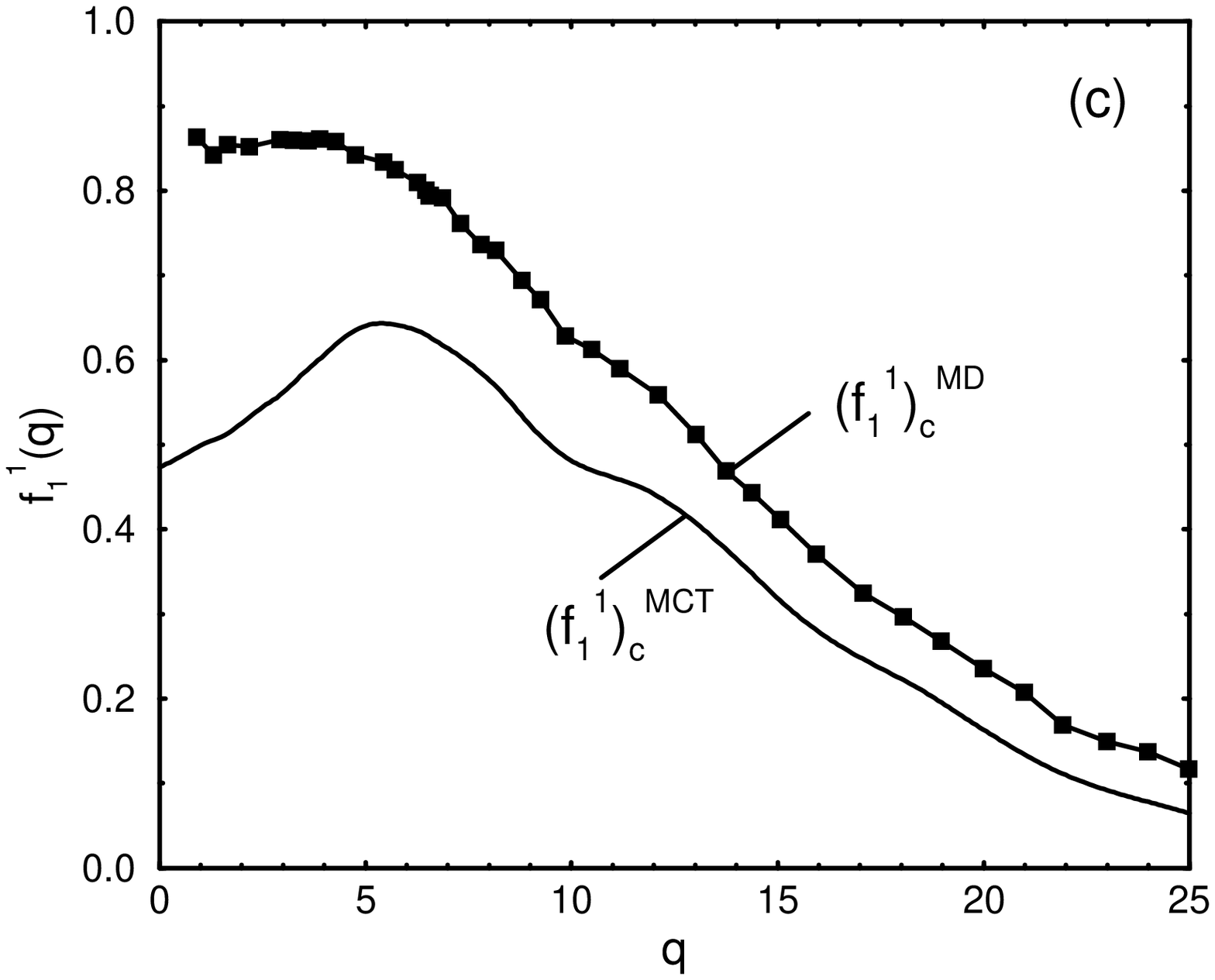,width=80mm}
\psfig{figure=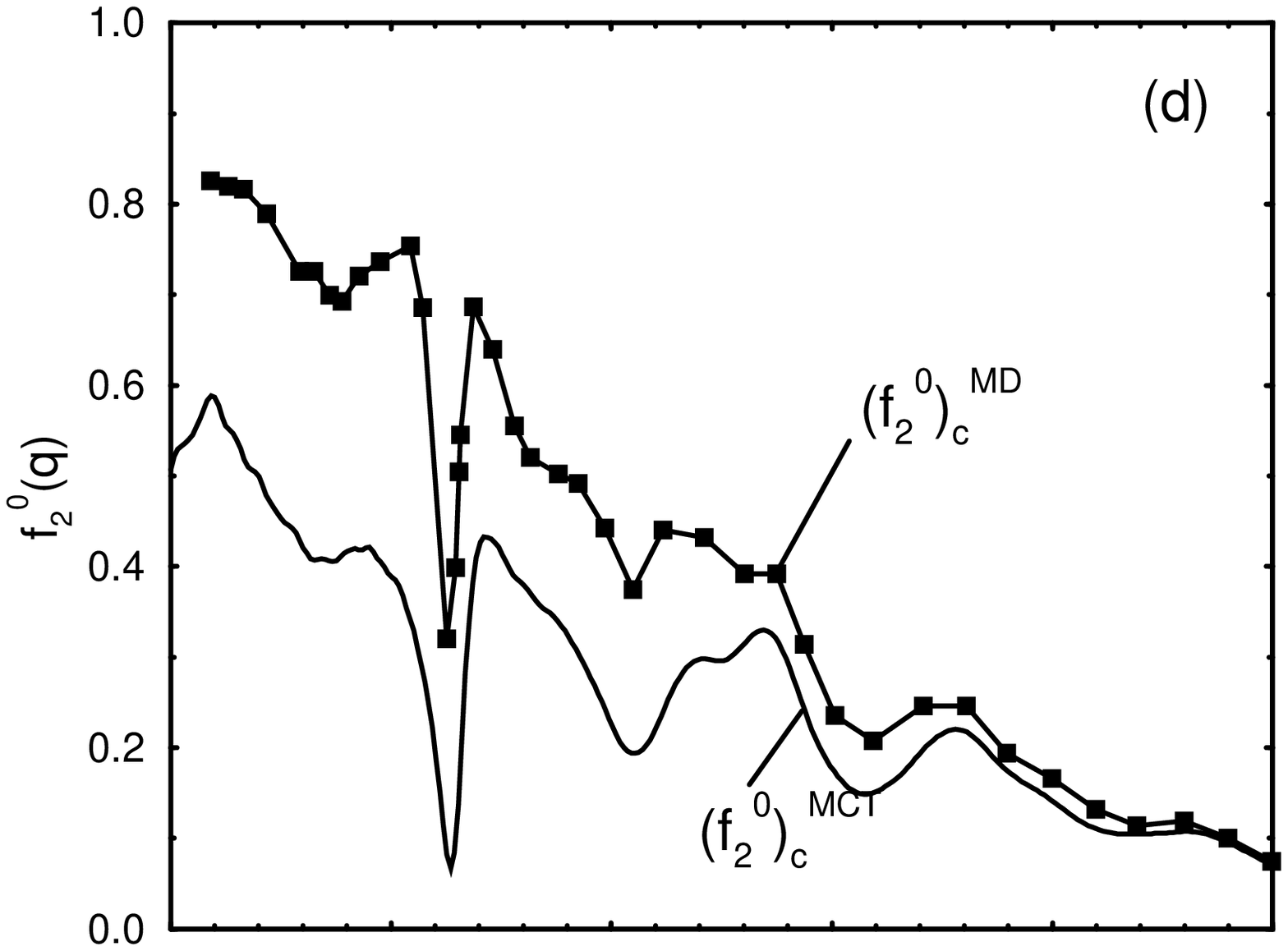,width=80mm} }
\centerline{\psfig{figure=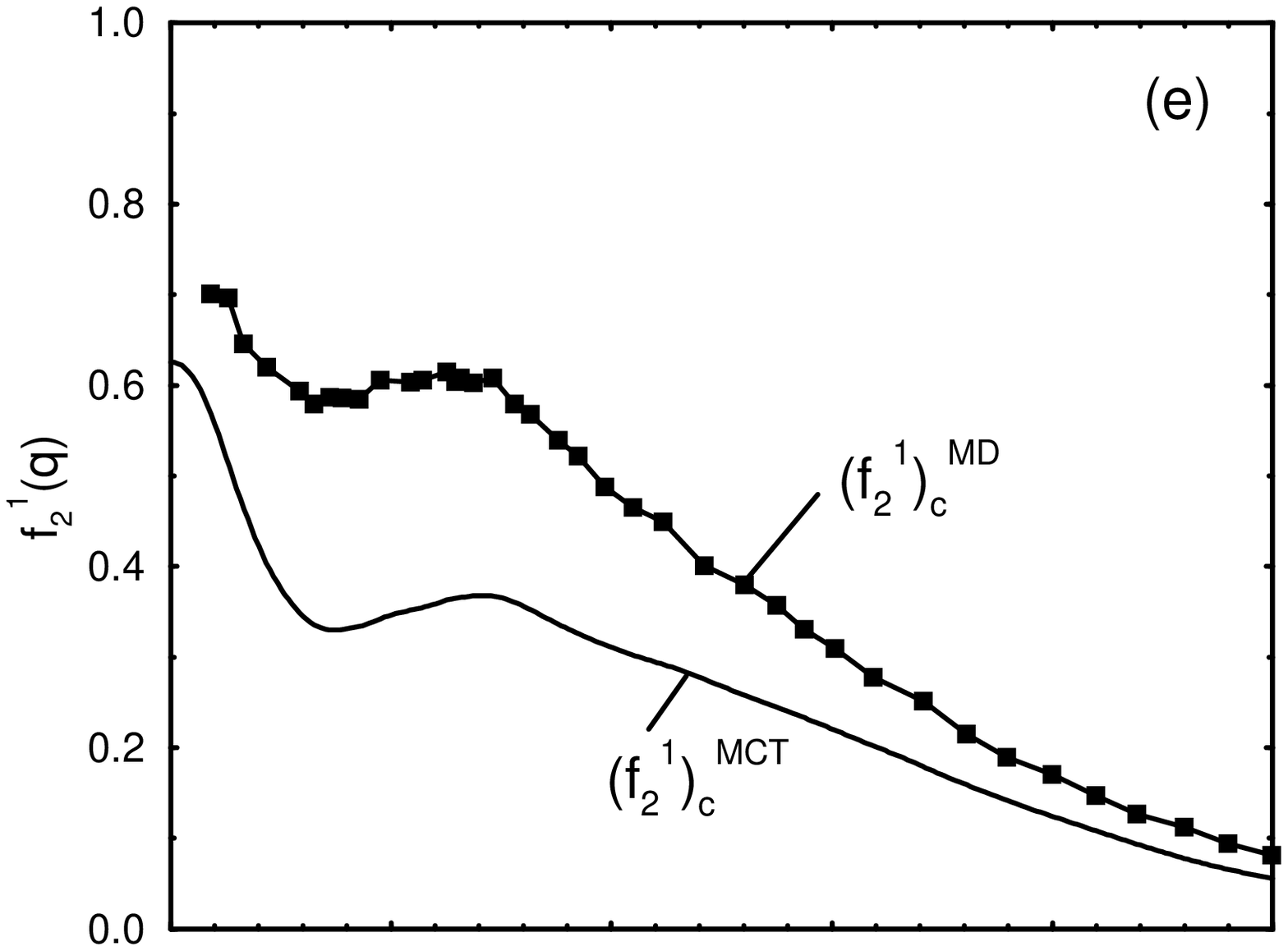,width=80mm}
\psfig{figure=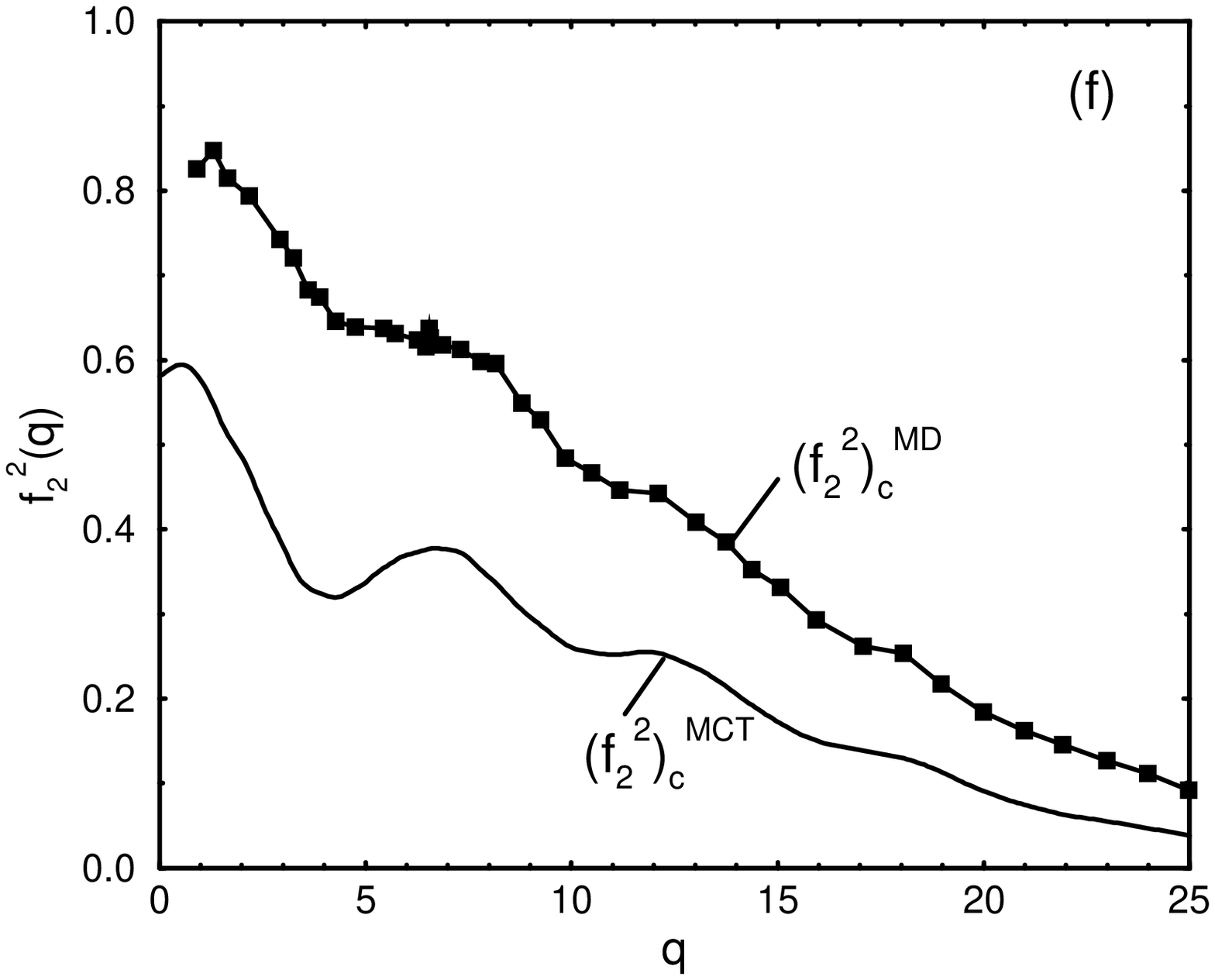,width=80mm} }
\caption[]{
q--dependence of the critical nonergodicity parameters
$(f_l^m)_c^{MD}$ from the MD--simulation and of
$(f_l^m)^{MCT}$ for $T \cong 0.310$: (a) $l = m = 0$, (b) $l = 
1, m = 0$, (c) $l = 1, m = 1$, (d) $l = 2, m = 0$, (e) $l = 2, m = 1$
and (f) $l = 2, m = 2$ . }
\end{figure}

\begin{figure}[htb]
\centerline{\psfig{figure=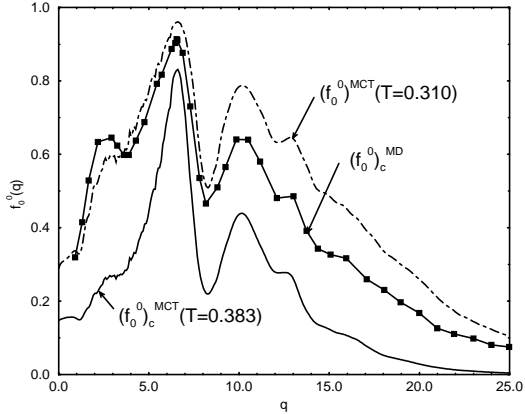,width=80mm,angle=270}} 
\caption[]{
Comparison of 
$(f_0^0)_c^{MD}$, $(f_0^0)^{MCT}(T_c^{l=0} \cong 0.383)$ and
$(f_0^0)^{MCT}(T \cong 0.310)$.} 
\end{figure}

\begin{multicols}{2}

Finally we compare
$\left(f_0^0\right)_c^{MCT}$ at $T = T_c^{l=0} = 0.383$, i.e. the
critical NEP, with that at $T \approx 0.310$. The result, including
$\left(f_0^0\right)_c^{MD}$ 
is given in Figure 3.
Now, we also find that $\left(f_0^0\right)_c^{MCT} <
\left(f_0^0\right)_c^{MD}$. That $\left(f_0^0\right)_c^{MCT} <
\left(f_0^0\right)^{MCT}(T \cong 0.310)$ is trivial, because $f_l^m$
increases monotonously with decreasing temperature.

\section{Summary and Conclusions}
\label{sec5}

For the first time, we have compared numerical results for a molecular
liquid of diatomic, rigid
molecules~\cite{kammererI,kammererII,kammererIII} with the
corresponding results from {\em molecular} MCT for linear molecules,
which was worked out recently~\cite{schilling97,theis97}. This
comparison was done for the nonergodicity parameters $f_l^m(q)$.
First of all, we find that MCT does not yield an ideal glass
transition in the temperature range of the simulation. This is quite
similar to what has been found for the MD--simulation of
water~\cite{sciortino97,fabbian97}, but is in variance with the behavior
for simple Lennard--Jones liquids~\cite{bengtzelius86,nauroth97} and
hard spheres~\cite{gotze91,gotze92}. For the hard sphere system this
means that the critical density $n_c$ from MCT is {\em less} that the
experimental value. This difference between simple and molecular
liquids probably reveals the role of ODOF and their coupling to TDOF.
This coupling is underestimated by both diagonalization approximations
we used. With these approximations the ideal glass transition occurs
below $T_c^{MD}$, only. The existence of two different transition
temperatures for $l_{co}=1$ and $l_{co}=2$ we believe is an artifact 
of the
diagonalization approximation (with respect to $l$). Nevertheless, the
q--dependence of $\left(f_l^m\right)_c^{MD}$ is reasonably well
reproduced by the corresponding MCT--result.
Whether the hump in
$\left(f_0^0\right)_c^{MCT}$ at $q \approx 3$, which relates to the
shoulder of $\left(f_0^0\right)_c^{MD}$ at the same q--value is
genuine can not presently be decided. But it is interesting that a
similar hump in the critical NEP has recently
been found for the {\em molecular} glass former
orthoterphenyl~\cite{tolle97}. Whether its existence in the
nonergodicity parameter $\left(f_0^0(q)\right)_c$ for the center of
mass motion is an indirect influence of the ODOF or not remains an
open question. For OTP this prepeak may be related to an {\em
intramolecular} property, not related to ODOF~\cite{tolle}.

To obtain the MCT--results presented in this paper already a
considerable numerical effort was necessary. Nevertheless it will be
important to solve the MCT--equations without any diagonalization
approximation, which we believe guarantees (for our molecules) the
existence of a {\em single} transition temperature
$T_c^{MCT}$. In this respect we also would like to
mention that, e.g. the {\em nondiagonal}, static correlator $S_{2
0}^0(q)$ is of the same order as $S_{0 0}^0(q)$~\cite{kammererIII},
i.e. it should not be neglected. Without assuming diagonality with
respect to $l$, more contributions to the translational-rotational
coupling will be taken into account which we expect to yield a higher
value for $T_c^{MCT}$. If in that case and for $l_{co}=2$ it would be
still $T_c^{MCT} < T_c^{MD}$, this would indicate that also
correlators with $l > 2$ have to be taken into account, as it was done
for the single dumb--bell in an isotropic liquid of hard spheres,
however, using SDA~\cite{franosch97}.
In ref.~\cite{franosch97} it has been found that a one percent accuracy
of $(f_l^m)_c$ requires to take into account correlators up to $l+2$.
But on the other hand our investigation shows that the SDA may
introduce an error of 5--10 \%. Therefore, in order to get reliable
results for the NEP, e.g. $(f_2^m)_c$ one has to remove both the
diagonalization approximation and the restriction to $l \le 2$.
\newline
\vspace{0.5cm}
\noindent 
\newline Acknowledgment:
We thank Wolfgang G\"otze and Arnulf Latz for critical 
comments on this manuscript
and also gratefully acknowledge the financial support by the
Sonderforschungsbereisch 262.

\end{multicols}

\end{document}